\documentclass[prl,twocolumn,showpacs,amsmath,amssymb]{revtex4}
\usepackage{amssymb}
\usepackage[dvips]{graphicx}
\usepackage[usenames]{color}
\usepackage{srcltx}
\usepackage{ulem}

\begin{document}
\title{Dynamics of resistive state in thin superconducting channels}
\author{ V. V. Baranov$^{1}$, A. G. Balanov$^{2}$, and V. V. Kabanov$^{1}$}
\affiliation{$^{1}$Department for Complex Matter, Jozef Stefan Institute, Jamova 39, 1001 Ljubljana, Slovenia;
$^{2}$Department of Physics, Loughborough University, LE11 3TU Loughborough, United Kingdom}
\date{\today}
\begin{abstract}

We theoretically study how the dynamics of the resistive state in narrow superconducting channels shunted by an external resistor depends on channel's length $L$, the applied current $j$, and parameter $u$ characterizing the penetration depth of the electric field in the nonequilibrium superconductors. We show that changing $u$ dramatically affects both the behaviour of the current-voltage characteristics of the superconducting channels and the dynamics of their order parameter. Previously, it was demonstrated that when $u$ is less than the critical value $u_{c1}$, which does not depend on $L$, the phase slip centers appear simultaneously at different spots of the channel. Herewith, for $u>u_{c1}$ these centres arise consecutively at the same place. In our work we demonstrate that there is another critical value for $u$. Actually, if $u$ does not exceed a certain value $u_{c2}$, which depends on $L$, the current-voltage characteristic exhibits the step-like behaviour. However, for  $u>u_{c2}$ it becomes hysteretic. In this case, with increase of $j$ the steady state, which corresponds to the time independent distribution of the order parameter along the channel, losses its stability at switching current value $j_{sw}$, and time periodic oscillations of both the order parameter and electric filed occur in the channel. As $j$ sweeps down, the periodic dynamics ceases at certain retrapping current value $j_r<j_{sw}$. Shunting the channel by a resistor increases the value of $j_r$, while $j_{sw}$ remains unchanged. Thus, for some high enough conductivity of the shunt $j_r$ and $j_{sw}$ eventually coincide, and the hysteretic loop disappears. We reveal dynamical regimes involved in the hysteresis, and discuss the bifurcation transitions between them.
\end{abstract}
\maketitle

\section{Introduction}

Understanding of the mechanisms responsible for destruction of the superconducting state is one of the most fundamental problems in physics of superconductors. It is known now that at the critical current a superconducting sample generally does not simply enter the fully normal state. The building up of an intermediate state structure \cite{skocpol}, flux flow \cite{anderson}, and heating effects \cite{pekker, li} lead to complicated transition behaviour \cite{tidecks}.

A number of very interesting phenomena have been observed for quasi-low-dimensional superconductors (see Ref. \cite{dmitrenko} for a review). The measurements of the current-voltage characteristics (CVCs) of such superconductors revealed a large transition region between the first occurrence of non-zero voltage and the completely normal state \cite{sivakov, zhuravel}. Within this transition region the voltage increases in a series of regular voltage jumps. The experimental and theoretical studies showed that every voltage jump is associated with the appearance of a localized phase-slip center (PSC) \cite{IvlevKopnin, dmitriev} in narrow superconducting channels or a phase-slip line (PSL) \cite{sivakov, zhuravel, dmit_zol} in  wide superconducting films. Moreover, the electric properties of a superconductor between the superconducting and fully normal state is determined by a system of interacting PSCs \cite{skocpol}. Their dynamics is still poorly understood, and at the moment is of a great research interest \cite{lu-dac, vod_peet,ours,kim,rubinstein}.

Previously it was found out that the parameter $u=\tau_{\psi}/\tau_{\theta}$, where $\tau_{\psi}$ is the relaxation time of the amplitude of the order parameter (OP) and $\tau_{\theta}$ is the relaxation time of the phase of the OP, governs the dynamics of the PSCs in a narrow superconducting ring \cite{lu-dac, vod_peet}. With this, the superconducting channel can  demonstrate two dynamical regimes. If $u<1$, then the phase of the OP relaxes faster than the amplitude, and PSCs occurs more or less simultaneously at different regions of the channel (see Fig. 3b of Ref. \cite{lu-dac}). In the case of $u>1$ the PSCs appear consecutively at the same spot of the channel (see Fig. 3a of Ref. \cite{lu-dac}). In Ref. \cite{ours} it was shown that the steps in the CVC  can be associated with bifurcations of either steady or oscillatory dynamics of OP. These bifurcations can substantially complicate dynamics of the OP and eventually lead to appearance of such sophisticated phenomena as multistability and chaos \cite{ours}. Thus, the appearance of multiple PSCs can be explained in terms of bifurcation phenomena. For example, Refs. \cite{ours, kim, rubinstein} report that the period-doubling bifurcation can be involved in formation and evolution of multiple PSCs in the spatio-temporal dynamics of OP.

Another interesting property of narrow superconducting channels is that their CVCs are able to demonstrate quite wide hysteresis, which is found to arise from Joule heating and the strong temperature dependence of the resistance of the wire \cite{pekker, tinkham, bezryadin}. Shunting of the channel by a resistor narrows this hysteresis, and for small enough resistance the hysteresis disappears \cite{bezryadin}.

It was suggested that shunting is able to influence the phase-slip events, and hence can be used for control of dissipation and coherence in nanowires \cite{bezryadin}. However, the dynamical mechanisms involved in the formation of the hysteresis are still not clear, and the ways of how shunting drives the phase-slip events are not fully apprehended. In this paper we study the dynamics of the resistive state in a narrow superconducting channel shunted by an external resistor and explore the bifurcation mechanisms involved in the hysteresis of CVCs and also responsible for formation and evolution of PSCs.

The article is organized as follows. In Sec. II we introduce the model, which we used to study the properties of the superconducting channels. In Sec. III we discuss the behaviour of the resistive state of the unshunted channels for different values of the parameter $u$, and reveal the bifurcation mechanisms responsible for the destruction of the periodic dynamics at the retrapping current. We study the superconducting channels shunted with an external resistance in Sec. IV. Finally, we summarize the results in the Sec. V.

\section{Formulation of the problem}

In the present paper we study the properties of a superconducting channel of the length $L$ by simulation of the dimensionless time-dependent Ginzburg-Landau equations (TDGLEs) \cite{gor-kop}:
\begin{equation}
u\Big(\frac{\partial\psi}{\partial t}+i\phi \psi\Big) =\frac{\partial^{2}\psi}{\partial x^{2}}+\psi-\psi|\psi|^{2},\label{tdgl1}
\end{equation}
\begin{equation}
j=-\frac{\partial\phi}{\partial x}+\frac{1}{2i}\Big(\psi^{*}\frac{\partial\psi}{\partial x}-\psi\frac{\partial\psi^{*}}{\partial x}\Big).\label{curr1}
\end{equation}
Here $\psi=\rho \exp(i\theta)$ is the dimensionless complex OP, where $\rho$ is the amplitude and $\theta$ is the phase of the OP. The distance is measured in units of the coherence length $\xi$ and time is measured in units of the phase relaxation time $\tau_{\theta}=\frac{4\pi\lambda^{2}\sigma_{n} }{c^{2}}$, $\lambda$ is the penetration depth, $\sigma_{n}$ is the normal state conductivity, and $c$ is the speed of light. The electrostatic potential $\phi$ is measured in the units of $\phi_{0}/2\pi c\tau_{\theta}$, where $e$ is the electronic charge and $\hbar$ is the Planck constant, the dimensionless current density $j$ is defined in the units of $\phi_{0}c/8\pi^{2}\lambda^{2}\xi$. The penetration depth of electric field $l_{E}$ is an additional parameter, which characterizes the nonequilibrium superconductors. From Eqs. (\ref{tdgl1}, \ref{curr1}) it follows that $l_{E}=\xi/u^{1/2}$. Therefore, the parameter $u$ characterizes the penetration of the electric field in the nonequilibrium superconductors. Notably, the value of the parameter $u$ introduced in the previous section depends on the material of superconductor.

Let us discuss the range of applicability of the TDGLEs in more details. One of the first attempts to analyse the simplified version of the TDGLEs similar to Eqs. (\ref{tdgl1},\ref{curr1}) was discussed in Ref.  \cite{KramerBaratoff}. In this work TDGLEs with $u=12$ were used to describe the gapless case with large concentration of magnetic impurities $\tau_{s}T_{c} <<1$, where $\tau_{s}$ is the scattering time on magnetic impurities, and $T_c$ is the transition temperature \cite{Eliashberg, KopninKniga}. Note that in the gapless case the TDGLEs can be rigorously derived from the microscopic theory. In that respect they are exact and are valid in the broad range of temperatures. More realistic generalized TDGLEs were derived by Kramer and Watts-Tobin \cite{KramerWatts}. Their version of the TDGLEs with $u=5.79$ takes into account inelastic scattering by phonons and  are valid for dirty superconductors  with a finite energy gap $\tau_{im}T_{c} <<1$, where $\tau_{im}$ is impurity scattering time. In the the dirty limit the inelastic scattering time is large. It leads to the enhancement of the penetration depth of the electric field $l_E$. This process can be described by the generalized TDGLEs \cite{
IvlevKopninMaslova} assuming that $u$ is small enough ($u<1$). Note that the generalized TDGLEs have an additional time derivative term which can lead to different dynamics at $j>j_{c}$ \cite{KopninKniga}. However, as it is pointed out in Ref. \cite{watts-tobin} (Eq.(196) and Fig. 8, page 499), this term does not affect the dynamics of the solution at large currents, while in the vicinity of the critical current the influence of this term has quantitative character. Therefore, Eqs. (\ref{tdgl1}, \ref{curr1}) describe well the gapless superconductors with large concentrations of magnetic impurities as well as superconductors with the gap near $T_{c}$. On the other hand variation of the parameter $u$ allows us to describe qualitatively ordinary superconductors in the dirty limit in the large temperature range. Definitely TDGLEs can not be used to reveal the properties of the superconducting channels at $T\to0$. However, some qualitative predictions based on this model can be used to describe quantum phase slips in this limit \cite{bez_rev, arutyunov}.

In the present paper we model a thin superconducting channel connected at the edges to the bulk superconductors. In this case, the boundary conditions are $\rho(-L/2)=\rho(L/2)=1$ and $d\phi(-L/2)/dx=d\phi(L/2)/dx=0$, where the phase $\theta$ and modulus $\rho$ of the OP satisfy the relation $\psi=\rho\exp{(i\theta)}$. The absence of the electric field at the edges of the channel determines the gradient of phase $d\theta(-L/2)/dx=d\theta(L/2)/dx=j$.

\section{Gapless case vs. superconductor with a gap}

In this section we study the superconducting channels of the lengths $L=L_0$, and  $L=2L_0$, where $L_0/\xi=11.78$. Here we use the Ginzburg-Landau depairing current for a channel of a finite length $j_c(L)$ as the reference point defining $j_c(L_0)/j_c=1.016$ and  $j_c(2L_0)/j_c=1.002$, where $j_c=2/(3\sqrt{3})$ is the Ginzburg-Landau depairing current for an infinite channel.

The value of the parameter $u$ plays the crucial role in the process of creation of the PSCs. Similarly to the results of the previous research \cite{lu-dac, vod_peet}, we found out that critical value of the parameter $u$, $u_{c1}=1$, corresponds to the change of the dynamics of the PSCs. Actually, for $u>u_{c1}$ the phase of the OP $\theta$ relaxes faster then the amplitude $\rho$. It leads to the appearance of the PSCs at the same spot of the channel, one after another (Figs. 1a - 1c). Our simulations suggest that such behavior of the PSCs remains qualitatively the same for any studied channel's length, and any studied value of the applied current. For $u<u_{c1}$ the amplitude $\rho$ relaxes faster then the phase $\theta$. In this case the PSCs can emerge in different regions of the channel (Fig. 1d). Hence, one can distinguish three different ways for the appearance of the PSCs: 1. There is only one PSC and the amplitude of the OP reaches zero several times at the same location. This is the case of the consecutive PSCs; 2. There are more than one PSCs and the amplitude of the OP reaches zero at the same time at all locations; 3. There are more than one PSCs, but the amplitude of the OP reaches zero at a different time for each location. Hereafter we refer to the cases 2. and 3. as to the case of simultaneous PSCs.

Besides $u_{c1}$, we reveal another critical value of the parameter $u$, $u_{c2}(L)$, which in contrast to $u_{c1}$ depends on the length of the channel. The critical values, $u_{c1}$ and $u_{c2}$, divide the parameter plane $(L, u)$ into the regions corresponding to the different type of solutions of the TDGLEs (see Fig. 2). Dynamics of PSC  for $u<u_{c2},u_{c1}$ (domain SS in Fig.2) was partly discussed in Ref. \cite{ours}. In this case the first PSC appears in the center of the channel (see Fig. 1c), and CVC in the vicinity of $j=j_c(L)$ can be approximated as $V\propto \sqrt{(j-j_c(L))/j_c}$ (the dashed line in Fig. 3). Analysis in Ref. \cite{ours}  showed that the period of OP oscillations in the vicinity of $j=j_c(L)$ for $u\le u_{c1}$ is described by formula
\begin{equation}
T=\frac{\pi\sqrt{3}(u+2)}{2^{3/2}u}\bigl[(j-j_{c})/j_{c}\bigr]^{-1/2}.
\label{period}
\end{equation}
\begin{figure}
\includegraphics[width = 85mm, angle=0]{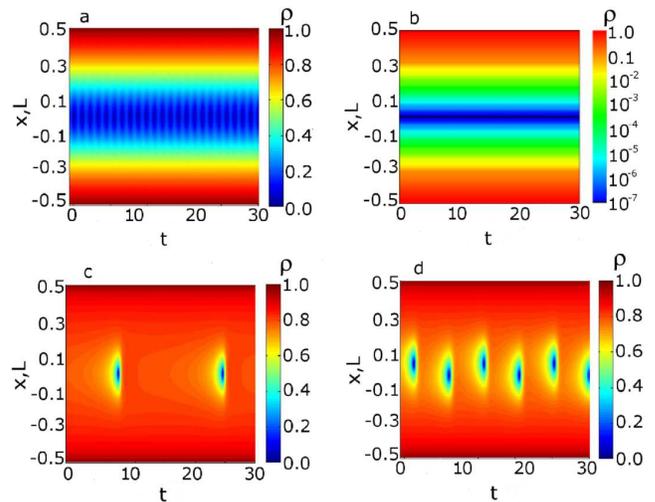}
\caption{The arrangement of the PSCs: a) in the vicinity of $j_{sw}$ for $u>u_{c1}$; b) for $j\gg j_{sw}$ and  $u\ge u_{c1}$; c) after the period-doubling bifurcation; d) in the vicinity of $j_c(L)$ for $u\le u_{c1}$ and in the vicinity of $j_r$ for $u> u_{c1}$.} \end{figure}
\begin{figure}
\includegraphics[width = 80mm, angle=0]{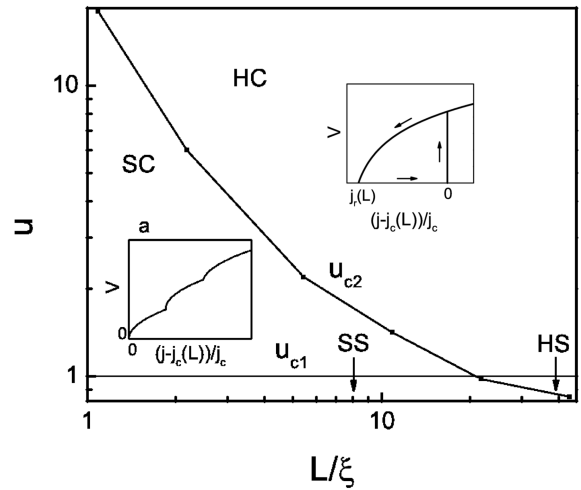}
\caption{Phase diagram showing regions of stability of different solutions of
the TDGL corresponding to the step like CVC for the consecutive (SC) and
simultaneous (SS) PSCs and the hysteretic one for the consecutive (HC) and
simultaneous (HS) PSCs. Insets represent the CVCs corresponding to the different
areas of the phase diagram.}
\end{figure}
Equation (\ref{period}) (solid and dashed lines in Fig.4b, which correspond to $u=0.5$ and $u=0.8$ respectively) demonstrates excellent agreement with the results of the numerical simulations (squares and triangles in Fig. 4b). The increase of the current leads to the period-doubling bifurcation, which implies a sudden increase of the period at $(j-j_c(L))/j_c=0.1$ for $L=L_0$ and $u=0.5$, $(j-j_c(L))/j_c=0.01$ for $L=2L_0$ and $u=0.5$, and $(j-j_c(L))/j_c=0.0046$ for $L=2L_0$ and $u=0.8$ (Fig. 4), and corresponds to the change of the spatial position of the PSCs: two adjacent PSCs appear symmetrically with respect to the center of the wire (Fig. 1d). Within the SS region in Fig. 2 it is also possible to find more complex solution of the TDGLEs, which are characterized by larger number of the PSCs \cite{ours}. Remarkably that the appearance of every new PSC induces a jump in the CVC as shown in Fig. 3.

\begin{figure}
\includegraphics[width = 70mm, angle=0]{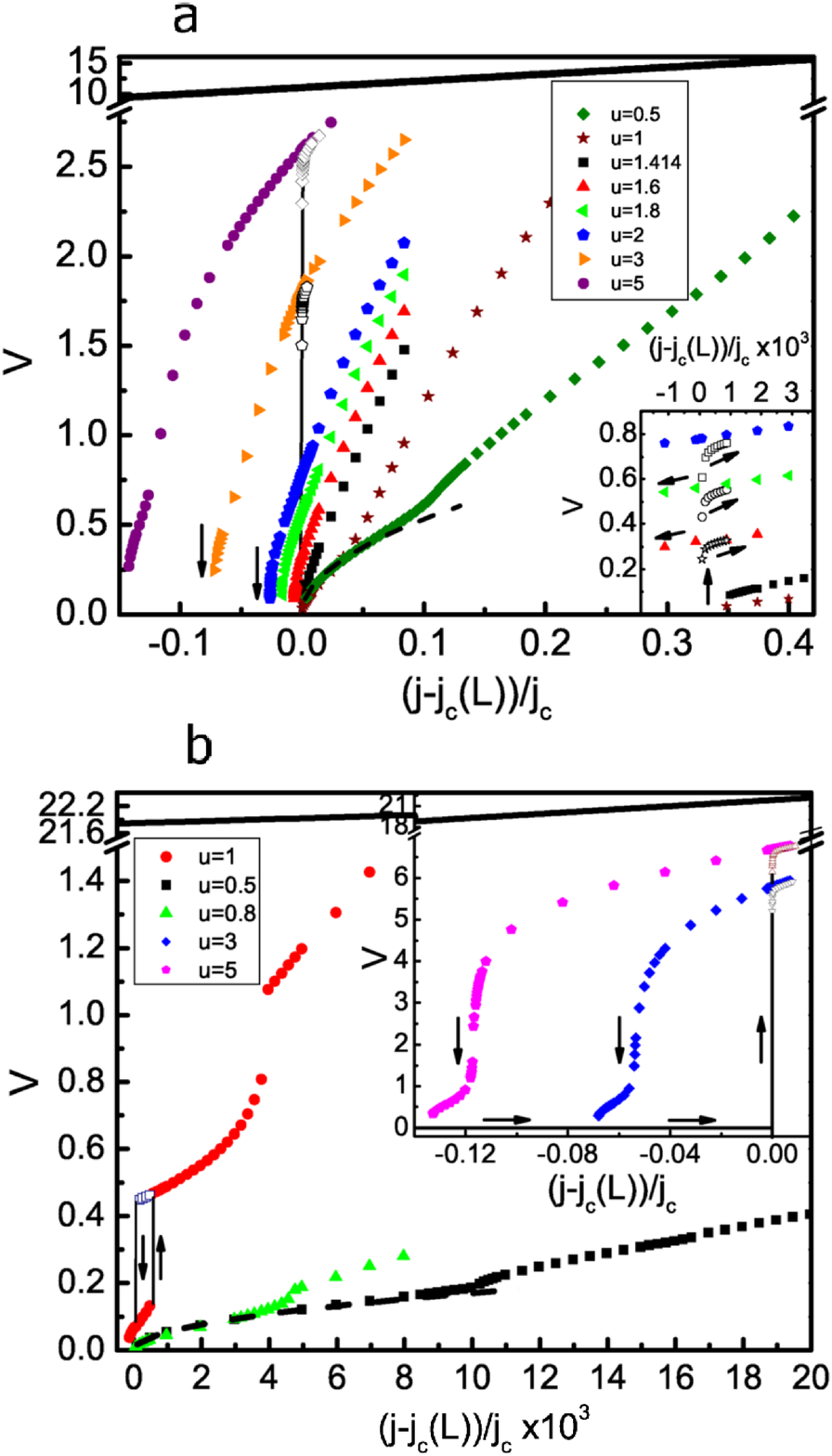}
\caption{The CVCs of the channels of the length a) $L=L_0$; b) $L=2L_0$ for different values of the parameter $u$. The heavy solid lines represent the CVC of the normal state. The dashed lines show the analytical formula ($V\propto\sqrt{(j-j_c(L)}$). Insets show a) hysteretic behaviour in the vicinity of $j_c(L)$; b) the CVC of the channels corresponding to the case $u>1$. The arrows illustrate hysteretic behaviour of the CVC.} \end{figure}

The HC regime in Fig. 2 corresponds to the qualitatively different behaviour of the CVCs. When $j<j_c(L)$, the steady-state solution is stable (see insets to Figs. 3a and 3b). However, the increase of $j$  induces the voltage drop along the superconducting channel. The critical current $j=j_{sw}$ corresponding to this bifurcation is called the switching current \cite{li} $j_c(L)=j_{sw}$. This switching is associated with the appearance of a single PSC in the center of the channel (Fig. 1a). Further increase of $j$ does not change the spacial position of the PSCs, but decreases  the period of the TDGLEs solution.  The OP is strongly suppressed in the small region in the center of the channel. The dimensionless amplitude of the OP in this area is  $\rho\sim10^{-7}$, while at the edges of the channel $\rho=1$ (Fig. 1b). The period of the solution tends to zero with increase of the current (see Fig. 4a and inset to Fig. 4b).

\begin{figure}
\includegraphics[width = 68mm, angle=0]{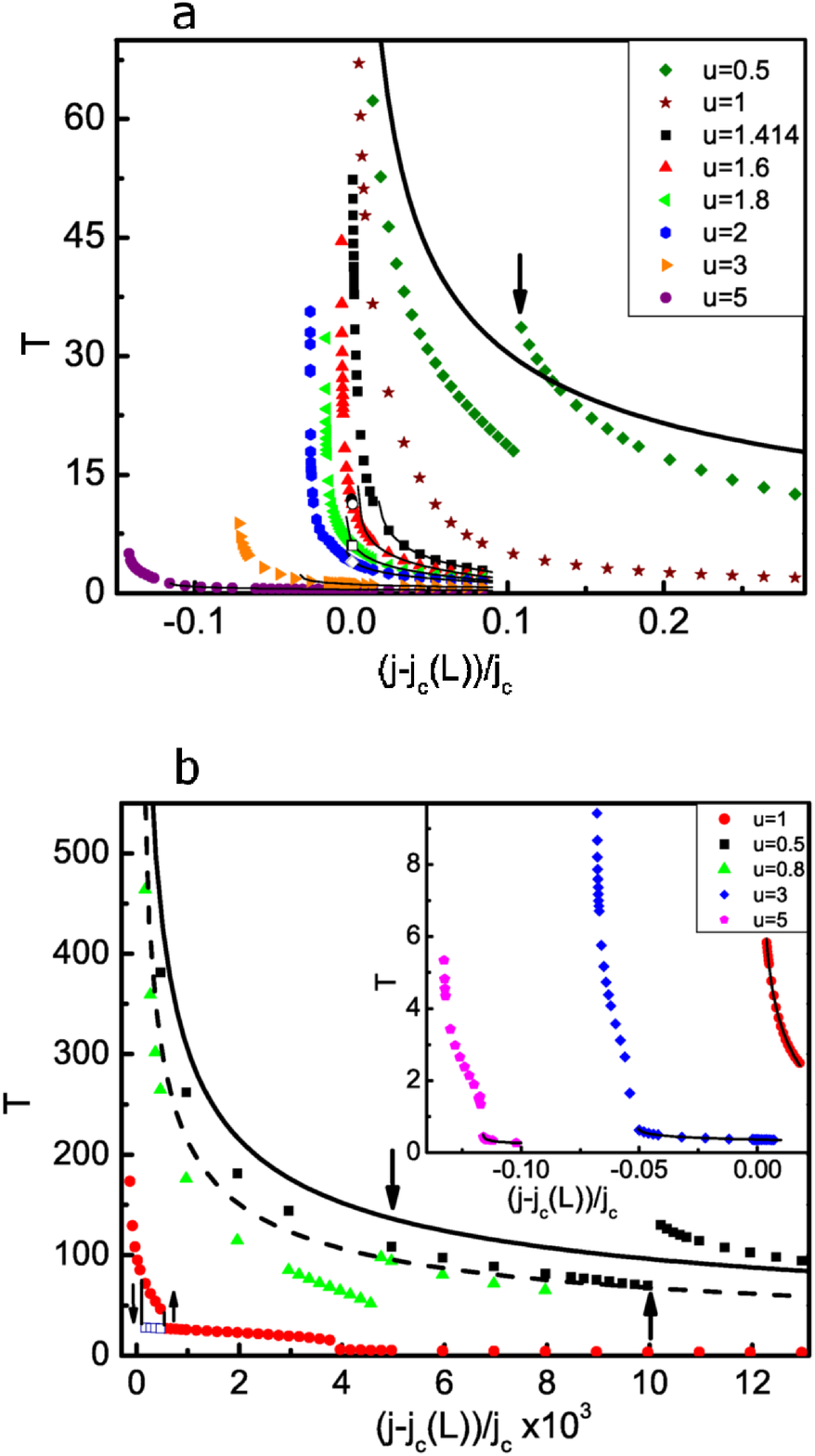}
\caption{The period of the solution as a function of the current of the channels of the length a) $L=L_0$; b) $L=2L_0$ for different values of the parameter $u$. Heavy solid and dashed lines represent the analytical formula (Eq. 3) for $u=0.5$ and $u=0.8$ respectively. Thin solid lines represent approximations by the logarithmic function (Eq. 4). The arrows illustrate the period-doubling bifurcation point.} \end{figure}

If now one decreases the current, the superconducting state restores for $j<j_{sw}$ indicating of the presence of a hysteresis. The critical current, $j=j_r$, below which the superconducting state reappears is known as the retrapping current \cite{li}. The CVCs of the channels in the HC regime has a pronounced tail in the vicinity of $j_r$ (see inset of Fig. 3b). The finite effective resistance appears in the range of $j<j_r$, which induces the change of the slope of the CVC. Similar effects were reported in \cite{vodolazov} for capacitively shunted Josephson junctions. However, in contrast to the results of Ref. \cite{vodolazov}, the system under consideration demonstrates a hysteretic type of CVCs. For $j<j_r$ the PSC appears in the center of the channel with significantly lower value of the voltage and higher value of the period comparing to the case of $j>j_r$ (Fig. 1c). Such behaviour of the PSCs is similar to the case of $u\le u_{c1}$, when we observe one PSC in the middle of the channel in the vicinity of $j_c(L)$.

The effective resistance of the channel grows with increase of the value of the parameter $u$ (Fig. 3). However, for a large range of $j$ it still remains much lower than the effective resistance of the normal state (solid line on Fig. 3). Nevertheless,  with increase of the current the CVC of the superconducting channel in the resistive state approaches the CVC of the normal state. These results are in agreement with the experimental results \cite{skocpol, zhuravel, patel}, where the CVCs of the superconducting channels at various temperatures have been investigated.

\begin{figure}
\includegraphics[width = 90mm, angle=0]{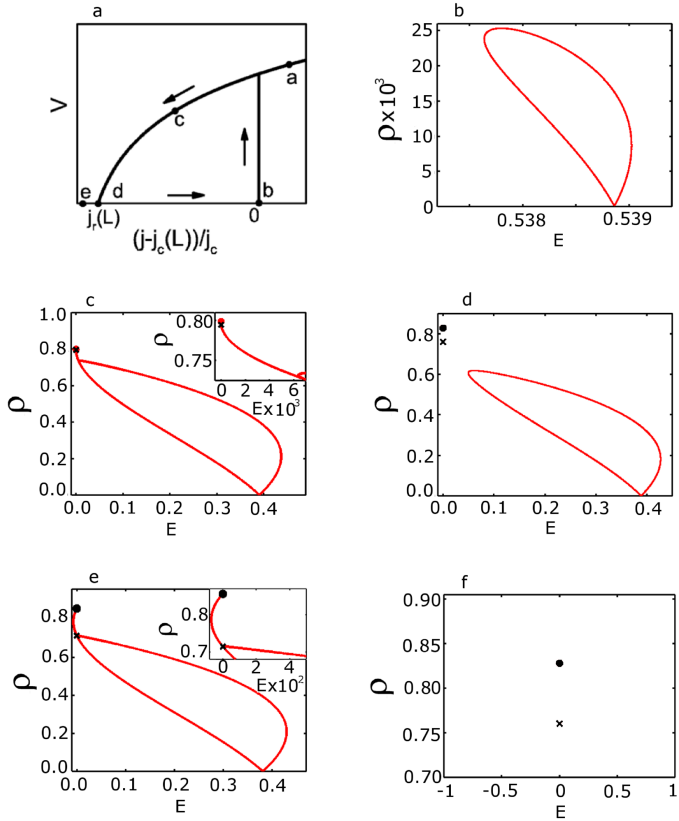}
\caption{ a) Hysteretic CVC corresponding to the HC domain on Fig. 2 obtained for parameter values $L=L_0, \quad u=2$; b)-f) Projections of the phase trajectories to $[\rho(0), E(0)]$ plane illustrating the structural changes that occur along the point sequence $a, b, c, d,$ and $e$ in inset a) corresponding  to the value of the current $(j-j_c(L))/j_c=0.05, 0, -0.016, -0.026, -0.03$ respectively. Full circle and cross represent the stable and unstable steady-state solution of the TDGLEs respectively.} \end{figure}

The hysteretic behaviour of the CVCs in the HC regime (Fig. 5a) can be illustrated by the transition between different solutions of the TDGLEs. For $j>j_{sw}$ the TDGLEs has only one periodic solution, which is represented in the phase space by a stable limit cycle (Fig. 5b), and which corresponds to a single branch in CVC (see Fig. 5a). With decrease of $j$,  at $j=j_{sw}$ a pair of steady states are born in the vicinity of the limit cycle as the result of saddle-node homoclinic bifurcation (Fig. 5c). With this, one of these steady states is stable, whereas another one is unstable. The coexistence of two stable solutions (bistability), a steady state and the limit cycle, for $j<j_{sw}$ stipulates two branches in the hysteretic CVC illustrated in Fig. 2b and Fig. 5a. With further decrease of $j$ the limit cycle approaches to the unstable steady state (Fig. 5d), and at $j=j_r$ they collide and disappear, suggesting a homoclinic bifurcation (Fig. 5e) \cite{Kuznetsov}. Remarkably, that in the vicinity of this bifurcation the period of the limit cycle can be approximated by the logarithmic law (thin solid lines on Fig. 4a and inset to Fig. 4b), similarly to the result obtained in Ref. \cite{IvlevKopnin}:
\begin{equation}
T=a-b\ln[(j-j_r(L))/j_c].
\label{period_sw}
\end{equation}
For smaller current $j<j_r$ the only limit solution of the system is the stable steady state, which corresponds to a single branch in the CVC in Fig. 2b and Fig. 5a. Thus, the destruction of the periodic solution at the retrapping current is associated with a homoclinic bifurcation \cite{Kuznetsov}. Note that the similar way of disappearance of OP oscillations was described in \cite{tidecks,michotte}. However, a particular instability leading to the destruction of the periodic solution was not revealed.

\begin{figure}
\includegraphics[width = 100mm, angle=0]{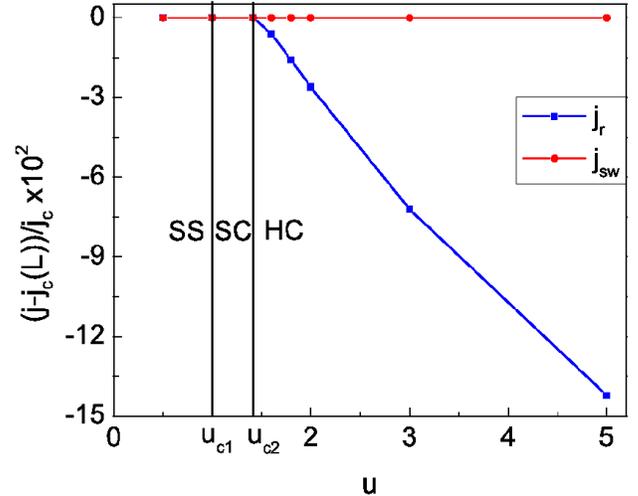}
\caption{The critical currents as a function of the parameter $u$ for the channel of length $L=L_0$. Squares and circles represent retrapping and switching currents respectively. } \end{figure}

The length of the superconducting channel $L$ strongly affects formation of PSCs, and correspondingly CVCs. For example, in the case of the longer channels $L\ge2L_0$ the value of $u_{c2}$ drops below $u_{c1}$ (Fig. 2). It leads to the unusual behaviour of the CVCs (see graph for $u=1$ in Fig. 3b) for $u_{c2}<u\le u_{c1}$ (HS domain  in Fig. 2) comparing with the CVCs corresponding to the other domains. Such unusual shape of CVCs can be understood in terms of evolution of PSC in the case of $L=2L_0$, $u=1$ (full circles in Fig. 3b). For these parameter values a single PSC (Fig. 1c) appears in the center of the channel in the vicinity of $j_{sw}$. Then, at the critical current corresponding to the period-doubling bifurcation, two PSCs appear symmetrically with respect to the center of the channel (Fig. 1d). The CVC and the period of the solution behave hysteretically in the vicinity of the period-doubling bifurcation. Then, at higher currents, the system switches to the HC regime, where a single PSC (Fig. 1b) appears in the center of the channel with the period lower than one of PSC in the vicinity of $j_c(L)$. Since the value of $u_{c1}$ in this case is close to $u_{c2}=0.99$, the value of the switching current ($j_{sw}=1.002$) is quite close to the value of the retrapping current ($j_r=1.0019$).

In the case of the shorter channels $L<2L_0$ there is another characteristic range of $u_{c1}<u<u_{c2}$ (domain SC in Fig. 2). For these values of $u$ the PSCs always appear in the center of the channel. Note that  in this case the CVC of the channel also possesses the step-like behaviour (stars on Fig. 3a). However, the change of the slope of CVC is not accompanied by creation of a new PSC. The period of the solution also remains unchanged (stars on Fig. 4a). Such behaviour of the CVC is similar to one in the HC regime, where a tail of CVC appears for $j<j_r$. However, in the former case the change of the slope of the CVC is more smooth. The absence of the period doubling bifurcation for $u_{c1}<u<u_{c2}$ is in agreement with the previous results \cite{lu-dac}, where it was shown that for $u>u_{c1}$ the PSCs appear consecutively.

The values of the switching $j_{sw}$ and retrapping $j_r$ currents are equal in the SS and SC regime (see Fig. 6). As it has been shown above, exceeding the critical value $u_{c2}$ leads to the hysteretic behaviour of the CVC. It corresponds to the appearance of the difference between the critical currents (domain HC on Fig. 6). Increase of the length of the channel causes decrease of the critical value $u_{c2}$. When $u_{c2}$ becomes smaller than $u_{c1}$, domain SC disappears. It corresponds to the appearance of the HS domain and shift of the SS domain to the region of small $u$, $u<u_{c1}$.

As it is clearly seen from Fig. 2, $u_{c2}$  sharply increases with decrease of the length of the channel $L$. It leads to the step like behaviour of the CVC for the very narrow channels. As it has been shown before, the CVC in the vicinity of the Ginzburg-Landau depairing current shows the square root behaviour in this region of the parameters \cite{ours}. It is in the agreement with the earlier result of Aslamazov, and Larkin \cite{AslamazovLarkin}, who have considered the CVC of the short ($L\le\xi$) superconducting bridge.

\section{A SC channel with a shunt}

\begin{figure}
\includegraphics[width = 90mm, angle=0]{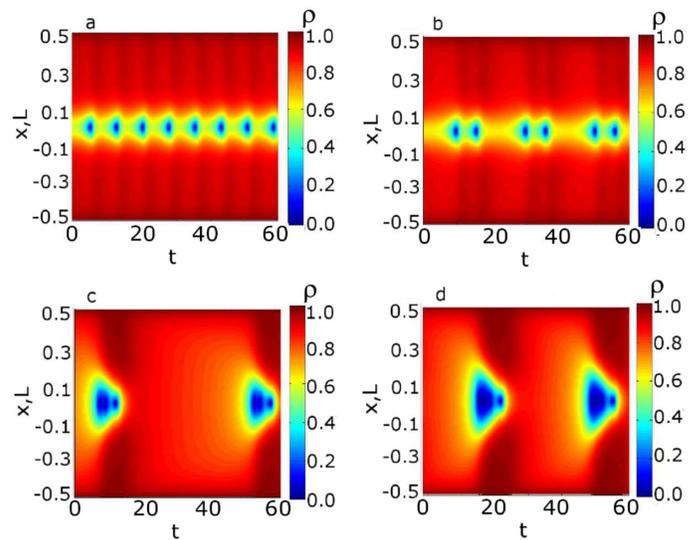}
\caption{The arrangement of the PSCs in the case of a shunted channel with: a) one; b) two; c) three; d) four PSCs per period.} \end{figure}
\begin{figure}
\includegraphics[width = 80mm, angle=0]{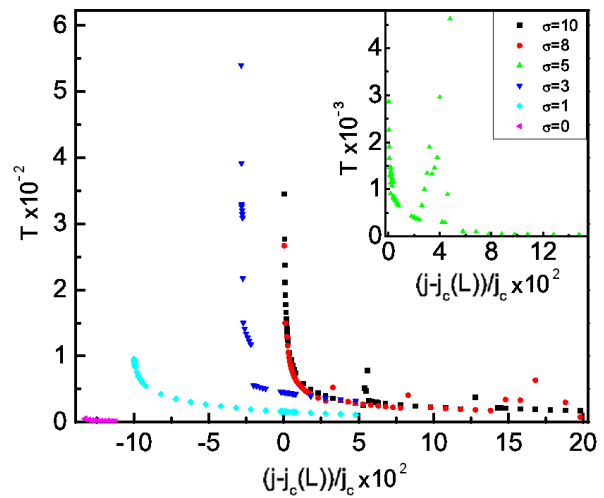}
\caption{The period of the solution as a function of the current of the channel of the length $L=2L_0$ for different values of conductivity. Inset represents the period of the solution for the value of conductivity $\sigma=5$.} \end{figure}
\begin{figure}
\includegraphics[width = 85mm, angle=0]{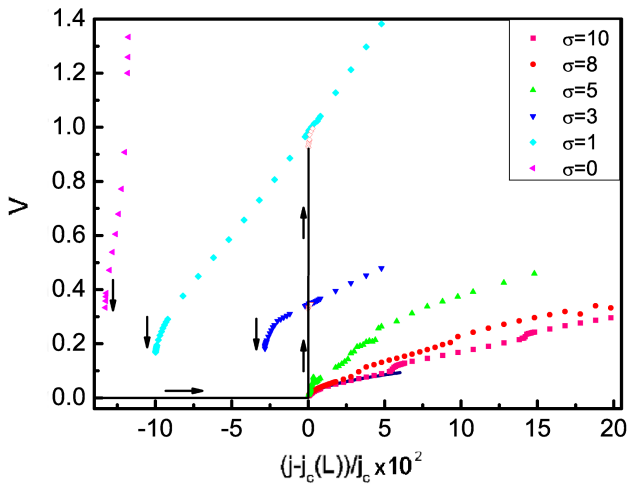}
\caption{The CVC of the channel of the length $L=2L_0$ for different values of conductivity of the shunt. The solid line shows the analytical formula ($V\propto\sqrt{(j-j_c(L)}$). The arrows represent hysteretic behaviour of the CVC.} \end{figure}

In this section we discuss the properties of a superconducting channel shunted with external resistors of different conductivity. Shunting the superconducting channel leads to decrease of the current flowing through the superconducting cannel. Therefore total current is defined as:
\begin{equation}
j=-\frac{\partial\phi}{\partial x}+\frac{1}{2i}\Big(\psi^{*}\frac{\partial\psi}{\partial x}-\psi\frac{\partial\psi^{*}}{\partial x}\Big)+\sigma_{sh}E\label{curr2}
\end{equation}
where $\sigma_{sh}$ is the conductivity of the shunt, and electric field in the shunt is defined as $E=-(\phi(L/2)-\phi(-L/2))/L$. In our study we fixed the value of $u=5$, which corresponds to a hysteretic CVC (see Fig. 3) for an unshunted channel (i.e. of a channel with $\sigma_{sh}=0$). For $u=5$ and $\sigma_{sh}=0$ a single PSC (Fig. 7a) appears for any values of $j$, however, increase of $\sigma_{sh}$ can dramatically affect the periodicity of the PSCs. For example, for $\sigma_{sh}=3$ a single PSC appears in the vicinity of the switching current during the up-sweep of $j$ (Fig. 7a). At the same time, down-sweep of $j$ induces a second PSC in the middle of the channel (Fig. 7b). Approaching the retrapping current $j_r$ increases the PSCs' number, and, finally, destroys the periodic solution. With this, every change of the periodicity of the PSC induces a discontinuity of the period of the solution of the TDGLEs (Fig. 8).

Growth of $\sigma_{sh}$ is accompanied by the increase of $j_r$, however, the value of $j_{sw}$ remains constant. Therefore, at certain critical value of $\sigma_{sh}=\sigma_c$ the value of $j_r$ and $j_{sw}$ coincides and the hysteresis disappears. Similar experimental results were obtained in Ref. \cite{bezryadin}. It was shown that shunting the superconducting channel with external resistors with different conductivities lead to different CVCs (see Fig. 2a of Ref. \cite{bezryadin}). If the conductivity of the shunt exceeds a certain critical value, then the CVC has the step-like structure. In the opposite case the CVCs have hysteretic character, in agreement with our calculations.  It should be noted however that the independence of $j_{sw}$ upon $\sigma_{sh}$ in our calculations is inconsistent with the earlier experimental results \cite{bezryadin}. This inconsistency may be associated with the neglecting of thermodynamic fluctuations in our model, whereas the values of $j_{sw}$ may be sensitive to the thermodynamic fluctuations.

Above the critical value of the conductivity, the dependencies of the voltage $V$ and the period of oscillations of OP $T$ on $j$ become quite complicated (see Fig. 9 and inset to Fig. 8). For $\sigma_{sh}\gtrsim\sigma_c$ the dependencies $V(j)$ and $T(j)$ demonstrate irregularities for low values of $j$. These irregularities are associated with appearance of new patterns in PSCs formation characterizing by different numbers of PSC. For high conductivity $\sigma_{sh}\gg\sigma_c$, the irregularities in the dependencies $V(j)$ and $T(j)$ shift towards larger values of $j$.

For example, for the channel of the length $2L_0$ the critical value of $\sigma_c=3.562$. Interesting that for $\sigma_{sh}>5$ a final voltage along the channel correspond to appearance of two (not one!)  PSCs in the center of the channel (Fig. 7b), which is similar to the results published in Ref. \cite{kim} Moreover, within a feasible range of $j$ we were unable to find a regime with only one PSC in the channel. Increase of $j$ upto 1.035 and 1.056 for $\sigma_{sh}=$ 8 and 10 respectively induces a third PSC in the middle of the channel (Fig. 7c), while further increases of $j$ upto 1.1 and 1.14 for $\sigma_{sh}=$ 8 and 10 respectively adds one more PSC (Fig. 7d). Remarkably, the PSCs for the above values of $j$ are rapidly follow one after another with relatively long time interval, which separates next group of PSCs. The number of PSCs per period increases quickly with the increase of $j$, which finally leads to the irregular behaviour of the period of OP (Fig. 8).

\section{Conclusion}

In conclusion, we revealed the dynamical mechanisms involved in the formation of the hysteresis, and showed how shunting drives the phase-slip events. We also studied the dynamics of the resistive state in a narrow superconducting channel shunted by an external resistor, and explored the bifurcation mechanisms which are responsible for formation and evolution of PSCs and which are also involved in the hysteresis of CVCs.

We have investigated the current-voltage characteristics of the superconducting channels of a different length. We have found two critical values of the parameter $u$. They induce the change of the dynamics of the solution of the TDGLEs. One of them is independent on the length of the channel ($u_{c1}=1$). It governs the periodicity and spatial position of the PSCs. They can be found in two regimes: simultaneous and consecutive ones. The second critical value of the parameter $u$ depends on the length of the channel ($u=u_{c2}(L)$). It changes the behaviour of the CVC from the step like to the hysteretic. As a result, the phase diagram showing the regions of stability of different solutions of the time-dependent Ginzburg-Landau equations can be divided to four regions corresponding to the step like current-voltage characteristics and the hysteretic one for the consecutive and simultaneous phase slip centers. The  current-voltage characteristics of the short $L\ll\xi$ superconducting channels possess the step like structure. The voltage in the vicinity of $j_c(L)$ is proportional to $\sqrt{j-j_c(L)}$. We also have revealed that the destruction of the periodic solution at the retrapping current corresponds to the homoclinic bifurcation. The results of our calculations show the switching between two metastable states in the area of the high currents regarding the switching current and $u>u_{c2}$. We have also investigated the superconducting channels shunted with an external resistor. It has been shown that the width of the hysteresis loop decreases with increase of the value of conductivity and disappears at a certain $\sigma_c$. In the transition from the hysteretic current-voltage characteristics to the step-like ones, some of them possess irregular behaviour. Different arrangements of the phase slip centers have been found for both the unshunted channels as well as for shunted ones. It has been shown that the change of the slope of a current voltage characteristics leads to the appearance of a new phase slip center.

\end{document}